\documentclass[conference]{IEEEtran}
\IEEEoverridecommandlockouts
\usepackage{cite}
\usepackage{amsmath,amssymb,amsfonts,amsthm}
\usepackage[linesnumbered,ruled,vlined]{algorithm2e}
\usepackage{graphicx}
\usepackage{tikz}
\usepackage{booktabs}
\usepackage{multirow}
\usepackage{xcolor}
\usepackage{url}

\newtheorem{theorem}{Theorem}
\newtheorem{lemma}{Lemma}
\newtheorem{corollary}{Corollary}
\newtheorem{definition}{Definition}
\newtheorem{observation}{Observation}

\newcommand{\degcbo}{\textsf{DegCbO}}
\newcommand{\ips}{\textsf{IPS}}
\newcommand{\oombea}{\textsf{ooMBEA}}
\newcommand{\imbea}{\textsf{iMBEA}}
\newcommand{\mbe}{MBE}
\newcommand{\cbo}{\textsf{CbO}}
\newcommand{\fcbo}{\textsf{FCbO}}
\newcommand{\lcm}{\textsf{LCM}}
\newcommand{\outv}{\mathrm{out}}
\newcommand{\inv}{\mathrm{in}}
\newcommand{\pos}{\mathrm{pos}}

\newcommand{\sat}{\mathrm{sat}}
\newcommand{\anc}[1]{G\langle #1\rangle}
\newcommand{\OO}{\mathcal{O}}
\newcommand{\Gm}{\Gamma}

\newcommand{\suf}{\mathrm{suf}}
\newcommand{\comm}{\mathrm{com}}
\SetKwInOut{Input}{Input}
\SetKwInOut{Output}{Output}
\SetKwFunction{Enum}{Enumerate}
\SetKwFunction{Setup}{Setup}
\SetKwFunction{PairSetup}{PairSetup}

\begin{document}

\title{Counting Maximal Bicliques in Large Sparse Bipartite Graphs\\ by Degeneracy-Anchored Enumeration\\ {\large [Technical Report]}}

\author{\IEEEauthorblockN{Lu Chen\IEEEauthorrefmark{1}, Chengfei Liu\IEEEauthorrefmark{1}, Rui Zhou\IEEEauthorrefmark{1}, Jiajie Xu\IEEEauthorrefmark{2}, Jianxin Li\IEEEauthorrefmark{3}}
\IEEEauthorblockA{\IEEEauthorrefmark{1}Swinburne University of Technology, Melbourne, Australia\\
\IEEEauthorrefmark{2}Soochow University, Suzhou, China\\
\IEEEauthorrefmark{3}Edith Cowan University, Perth, Australia\\
\IEEEauthorrefmark{1}\{luchen, cliu, rzhou\}@swin.edu.au, \IEEEauthorrefmark{2}xujj@suda.edu.cn, \IEEEauthorrefmark{3}jianxin.li@ecu.edu.au}}

\maketitle

\begin{abstract}
Given a bipartite graph, maximal biclique enumeration (\mbe) reports all the maximal bicliques and maximal biclique counting computes their number, both of which are fundamental for analysing bipartite data.
However, existing algorithms grow one side of a biclique vertex by vertex inside two-hop neighbourhoods.
As such, the size of their subproblems is governed by a quantity that is unbounded on hub-heavy graphs.
In this paper, we propose a novel algorithm, denoted as \degcbo, which anchors every maximal biclique at the first vertex $v$ of its vertex set in a degeneracy order, i.e., one of its sides is confined to the forward neighbourhood $\outv(v)$ with $|\outv(v)|\le d$, where $d$ is the degeneracy of the graph.
Instead of sweeping the $2^{|\outv(v)|}$ subsets, \degcbo\ enumerates only the maximal bicliques of the small anchored subgraph of $v$ with a canonicity test that rejects duplicates.
We show that maximality in the whole graph needs exactly two families of blockers beyond the anchored subgraph, both of which reduce to bitwise operations on precomputed masks.
The search tree has at most $n\cdot 2^{d}$ nodes and, since $d\le\sqrt{m}$ in a bipartite graph, the count is computed in $\OO^{*}(2^{\sqrt m})$ time, which matches the output lower bound of enumeration on the crown family.
Since the anchored subgraph must be materialised at every anchor, which costs more than the enumeration on hub-heavy graphs, a cost-aware sweep, an exact twin quotient and a corrected form of failure propagation are proposed, which together make the set-up output-sensitive.
Besides, we show that counting is not output-bound: a closed-form terminal counts star-shaped subtrees in constant time, and a pairwise form of the count is computable in $2^{o(n)}$ time by the recent sub-exponential algorithms for counting satisfying assignments.
In contrast, under the Satisfiability Hypothesis no algorithm counts in $2^{d^{1-\varepsilon}}$ time, i.e., the bound $n\cdot2^{d}$ is tight to the extent the hypothesis allows.
Extensive experiments are conducted on sixteen real graphs to justify that our proposed algorithm is faster than the state-of-the-art algorithm on every graph, by up to $35$ times.
\end{abstract}

\begin{IEEEkeywords}
maximal bicliques, counting, enumeration, bipartite graphs, degeneracy order
\end{IEEEkeywords}

\section{Introduction}\label{sec:intro}

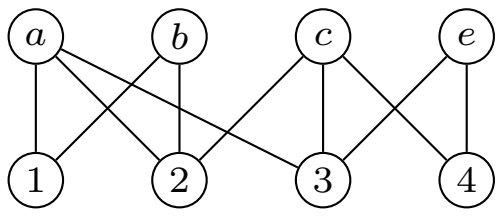
\begin{figure}[t]
\centering
\footnotesize
\begin{tabular}{@{}c@{\qquad}l@{}}
\begin{tikzpicture}[baseline=(current bounding box.center),x=0.95cm,y=0.95cm,
  every node/.style={circle,draw,inner sep=0pt,minimum size=3.6mm,font=\scriptsize}]
\foreach \p/\x in {a/0,b/1,c/2,e/3} \node (\p) at (\x,1) {$\p$};
\foreach \q/\x in {1/0,2/1,3/2,4/3} \node (r\q) at (\x,0) {$\q$};
\foreach \p/\q in {a/1,a/2,a/3,b/1,b/2,c/2,c/3,c/4,e/3,e/4} \draw (\p)--(r\q);
\end{tikzpicture}
&
\begin{tabular}{@{}l@{}}
$L=\{a,b,c,e\}$, $R=\{1,2,3,4\}$ \\
degeneracy order $\pi=(b,1,e,4,a,2,c,3)$ \\
$d=2$
\end{tabular}
\end{tabular}\\[2pt]
\textbf{(a) Bipartite graph $G$}\\[4pt]
\begin{tabular}{@{}l@{}}
\textbf{anchor $b$}: $C=\{1,2\}$; companions $a{:}11$, $c{:}01$, $b{:}11$; $\inv(b)=\emptyset$ \\
\quad root $\sat(\emptyset)=\{2\}$, companions $\{a,b,c\}$: report $(\{a,b,c\},\{2\})$ \\
\quad child $\{1,2\}$, companions $\{a,b\}$: report $(\{a,b\},\{1,2\})$ \\
\textbf{anchor $1$}: $C=\{a\}$ ($b\prec 1$); companions $2,3,1$, all ${:}1$; $\inv(1)=\{b\}$ \\
\quad root $\{a\}$, companions $\{1,2,3\}$; candidate-side ghost $b\not\sim 3$: report \\
\textbf{anchor $e$}: $C=\{3,4\}$; companions $a{:}10$, $c{:}11$, $e{:}11$ \\
\quad root $\{3\}$, companions $\{a,c,e\}$: report \\
\quad child $\{3,4\}$, companions $\{c,e\}$: report \\
\textbf{anchor $4$}: $C=\{c\}$ ($e\prec 4$); root $\{c\}$, companions $\{2,3,4\}$; \\
\quad candidate-side ghost $e\not\sim 2$: report $(\{c\},\{2,3,4\})$ \\
\textbf{anchor $a$}: $C=\{2,3\}$ ($1\prec a$); companions $c{:}11$, $a{:}11$; \\
\quad root $\{2,3\}$, companions $\{a,c\}$; candidate-side ghost $1\not\sim c$: report \\
\textbf{anchors $2,c,3$}: $C=\emptyset$ or every local maximal biclique is killed by a ghost \\
\end{tabular}\\[2pt]
\textbf{(b) Trace of \degcbo}
\caption{Running example}
\label{fig:example}
\end{figure}

A bipartite graph $G=(L,R,E)$ consists of two disjoint vertex sets $L$ and $R$ and an edge set $E\subseteq L\times R$.
A pair of vertex sets $(A\subseteq L, B\subseteq R)$ is a \emph{biclique} if every $(u,v)\in A\times B$ is an edge, and it is a \emph{maximal biclique} if it cannot be enlarged.
For instance, in Figure~\ref{fig:example}, $(\{a,b\},\{1\})$ is a biclique and $(\{a,b\},\{1,2\})$ is a maximal biclique.
Maximal biclique enumeration (\mbe) reports every maximal biclique of $G$ and maximal biclique counting computes their number, denoted as $\beta$.
Since the existing algorithms count by enumerating, the two problems are measured in the same way.

\noindent\textbf{Applications.}
Bicliques expose co-occurrence structure, e.g., short texts sharing common words~\cite{muhammad2016summarizing}, genes sharing behaviours in expression matrices~\cite{voggenreiter2012exact}, users visiting common web pages~\cite{kumar1999trawling}, and closed itemsets, where a maximal biclique of the transaction--item graph is exactly a closed frequent itemset~\cite{uno2004lcm}.
Since the constraints that make a biclique interesting are specific to an application, a common strategy is to enumerate the maximal bicliques and filter afterwards.
Besides, the maximum edge biclique and maximum balanced biclique algorithms~\cite{chen2021mbb,newMEB} are also built on \mbe\ machinery.
As such, \mbe\ has been studied extensively~\cite{zhang2014imbea,abidipivot,chen2022oombea,ips2026,das2019parmbe,gmbe2025}.
Real graphs with a few million edges hold $10^{10}$ to $10^{14}$ maximal bicliques.

\noindent\textbf{State-of-the-art.}
The practical algorithms are based on \imbea~\cite{zhang2014imbea}.
They work on one vertex set, say $R$, maintain the current biclique $(A,B)$ together with a candidate set $P\subseteq R$ and a set $Q\subseteq R$ of vertices that have been expanded previously, and branch by moving a vertex of $P$ into $B$ and intersecting $A$ with its neighbourhood.
\oombea~\cite{chen2022oombea} observes that for a sparse bipartite graph the vertices of $R$ reachable from a vertex $v$ through its neighbours, i.e., the two-hop neighbourhood $N_{2}(v)$, converge to a small set.
It introduces the unilateral order to bound the enumeration scope of every vertex by its unilateral coreness, and adds the batch-pivots technique.
The state-of-the-art algorithm~\cite{ips2026} follows the same two-hop approach.
All these algorithms share one property, i.e., the recursion grows $B$ one vertex at a time inside a two-hop neighbourhood.
As such, the depth of the recursion and the size of the subproblem at every node are governed by the largest two-hop neighbourhood that survives pruning, which reaches $3.7$ million on \textsf{LiveJournal}.
Therefore, the subproblems are unbounded by any linear-time graph parameter, and on hub-heavy graphs these algorithms do not finish.

\noindent\textbf{Uncharted opportunity for \mbe.}
Eppstein~\cite{eppstein1994arboricity} observes that in an acyclic orientation of bounded out-degree, every maximal biclique is \emph{anchored} at the first vertex $v$ of its vertex set, and one of its sides is a subset of $\outv(v)$.
With a degeneracy order, the out-degree bound is the degeneracy $d$ of $G$, which is computed exactly in $\OO(m)$ by min-degree peeling and is small on real bipartite graphs, e.g., $20$ on \textsf{youtube}, $18$ on \textsf{wikicat} and $108$ on the LiveJournal $108$-core, where the largest two-hop neighbourhood has $3.7$ million vertices.
However, the algorithm in~\cite{eppstein1994arboricity} sweeps all $2^{|\outv(v)|}$ subsets and is not competitive.
The opportunity is to visit only the subsets of $\outv(v)$ that can be the side of a maximal biclique, which are exactly the sides of the maximal bicliques of a small subgraph anchored at $v$.
They can be enumerated depth-first with the canonicity test of Close-by-One (\cbo)~\cite{kuznetsov1999cbo}, which rejects a biclique at every node but the one where it is first generated.

\noindent\textbf{Challenges.}
The first challenge is to bound the enumeration scope of a vertex by a graph parameter that is computed exactly in linear time and that is small on real graphs.
The unilateral coreness of \oombea\ is one such parameter, but it is still in the thousands on million-edge graphs.
The second challenge is that enumerating in the residual graph above $v$ breaks maximality, which must be re-established without the per-node set intersections over two-hop neighbourhoods that every \imbea-based algorithm performs.
The third challenge is created by anchoring itself.
When the enumeration is anchored at a vertex $v$, the subgraph in which the bicliques of $v$ are enumerated must be materialised from the adjacency lists of its forward neighbours.
Summed over all anchors, this costs $\sum_{c}\deg(c)\cdot|\inv(c)|=\OO(\Delta m)$, where $\Delta$ is the maximum degree.
On hub-heavy graphs this set-up scan, not the enumeration, dominates the running time, and it has to be made output-sensitive without touching the enumeration.
The fourth challenge is specific to counting.
Every existing algorithm counts by visiting a node per biclique, i.e., its time is at least $\beta$.
However, the count itself is a single number and, on the families that attain the worst case, a closed form.

\noindent\textbf{Our approach.}
In this paper, we propose a novel algorithm, denoted as \degcbo, i.e., degeneracy-anchored enumeration with \cbo\ canonicity.
The main ideas are as follows.

\noindent\underline{\textit{Anchored enumeration}}.
To address the first challenge, for every vertex $v$ in degeneracy order \degcbo\ builds a small \emph{anchored subgraph}.
Its candidate side is $\outv(v)$ and its other side consists of the residual vertices adjacent to some candidate, each of them a bitmask over the candidates.
\degcbo\ enumerates the maximal bicliques of that subgraph by their candidate sides.
To address the second challenge, we prove that maximality in the whole graph needs exactly two blocker families beyond the anchored subgraph: 1) \emph{candidate-side ghosts}, which are exactly the backward neighbours $\inv(v)$, and 2) \emph{anchor-side ghosts}, i.e., the earlier vertices adjacent to all of the current candidate side.
Both reduce to aggregate bitwise operations on precomputed masks.
The search tree has at most $n\cdot 2^{d}$ nodes.

\noindent\underline{\textit{Dealing with the set-up}}.
To address the third challenge, we propose a \emph{cost-aware sweep}, i.e., per anchor, we either scan the adjacency of every candidate or walk the co-occurrence lists of candidate pairs.
In the latter case, a vertex adjacent to exactly one candidate is never materialised.
This is because it can only belong to the biclique with that single candidate, whose saturation, canonicity and maximality are decided in $\OO(1)$ from counts.
Together with an exact \emph{twin quotient} of the input and a corrected form of \fcbo\ failure propagation, this removes the loss case.

\noindent\underline{\textit{Counting beyond enumeration}}.
To address the fourth challenge, we show two exact ways to count without visiting.
The first is a \emph{closed-form terminal} for star-shaped subtrees, which counts the crowns, the worst case of every enumeration algorithm, in $\OO(k^{2})$ nodes.
It also measures the structured share of $\beta$ on real graphs, of which the hardest instances have none.
The second is a pairwise form of $\beta$ under which the recent sub-exponential algorithms for counting satisfying assignments apply, giving a $2^{o(n)}$ bound.
In contrast, we prove that no algorithm counts in $2^{d^{1-\varepsilon}}$ time unless the Satisfiability Hypothesis fails.

Our principal contributions are summarised as follows.
\begin{enumerate}
\item We propose \degcbo, which anchors each maximal biclique at its degeneracy-first vertex and enumerates only the maximal bicliques of the anchored subgraph, with a search tree of at most $n\cdot2^{d}$ nodes.
We prove that $d\le\min\{\sqrt{m},\min(|L|,|R|)\}$ in a bipartite graph.
As such, \degcbo\ runs in $\OO^{*}(2^{\sqrt{m}})$, matches the worst-case output size, and is optimal up to an additive $d$ on the crown family (Section~\ref{sec:algo}).
\item We characterise maximality beyond the anchored subgraph, i.e., candidate-side extenders are exactly $\inv(v)$ and anchor-side extenders are exactly the earlier vertices covering the candidate side.
Both are necessary and both become bitwise operations (Section~\ref{sec:algo}).
\item We solve the set-up problem of anchored enumeration with the cost-aware sweep, the twin quotient and effective failure propagation, each of which is exact (Section~\ref{sec:setup}).
\item We show how to count without visiting, with the closed-form terminal and the $2^{o(n)}$ counting bound, and we prove a conditional lower bound of $2^{d^{1-\varepsilon}}$ for counting (Section~\ref{sec:count}).
Besides, the enumeration itself reduces to a single pass over bitmasks per node (Section~\ref{sec:impl}).
\item Extensive experiments are conducted on sixteen real graphs to justify that \degcbo\ is faster than the state-of-the-art algorithm on every graph, between $1.6$ and $35$ times.
It completes the largest instance reported in the serial literature, with $1.96\times10^{10}$ maximal bicliques, in just over an hour (Section~\ref{sec:exp}).
\end{enumerate}

\section{Problem Formulation}\label{sec:prelim}

\noindent\textbf{Bipartite graph.}
Let $G=(L,R,E)$ denote a bipartite graph with $n=|L|+|R|$ vertices and $m=|E|$ edges.
For instance, Figure~\ref{fig:example}(a) shows a bipartite graph with $L=\{a,b,c,e\}$ and $R=\{1,2,3,4\}$.
Given a vertex $u$, the neighbours and the degree of $u$ are denoted by $N(u)$ and $\deg(u)=|N(u)|$ respectively, and the maximum degree is denoted by $\Delta$.
Given a set of vertices $X\subseteq L$ or $X\subseteq R$, we use $\Gm(X)$ to denote the common neighbours of the vertices in $X$, i.e., $\Gm(X)=\bigcap_{u\in X}N(u)$.

\noindent\textbf{Maximal biclique.}
A pair of vertex sets $(A,B)$ with $A\subseteq L$ and $B\subseteq R$ is a biclique if $A\times B\subseteq E$.
If no vertex can be added to $A$ or to $B$, $(A,B)$ is a \emph{maximal biclique}.
Equivalently, $(A,B)$ is a maximal biclique if and only if $A=\Gm(B)$ and $B=\Gm(A)$.
The number of maximal bicliques of $G$ is denoted by $\beta$.

\noindent\textbf{Problem statement.}
Given a bipartite graph $G$, compute $\beta$, and in enumeration mode report every maximal biclique exactly once.
Following the state-of-the-art algorithm~\cite{ips2026}, we measure all algorithms in counting mode, in which a biclique is reported by incrementing a counter.
The output cost of writing bicliques is the same for every algorithm, and Section~\ref{sec:count} is about counting without reporting.

\noindent\textbf{Degeneracy order.}
The degeneracy $d$ of $G$ is the largest $k$ such that $G$ has a subgraph of minimum degree $k$.
Repeatedly removing a vertex of minimum degree produces a \emph{degeneracy order} $\pi$ in $\OO(m)$ time.
Orienting every edge from its earlier to its later endpoint gives an acyclic orientation in which every vertex has at most $d$ forward neighbours~\cite{eppstein2010els}.
We use $\pos(v)$ to denote the rank of $v$ in $\pi$.
The forward neighbours of $v$ are denoted by $\outv(v)$, i.e., $\outv(v)=\{w\in N(v):\pos(w)>\pos(v)\}$, and the backward neighbours of $v$ are denoted by $\inv(v)$, i.e., $\inv(v)=N(v)\setminus\outv(v)$.
As such, $|\outv(v)|\le d$.
The \emph{residual} at $v$, denoted by $G_{v}$, is the subgraph induced by $\{w:\pos(w)\ge\pos(v)\}$.
For instance, in Figure~\ref{fig:example}, $\outv(1)=\{a\}$ and $\inv(1)=\{b\}$ since $b$ appears before $1$ in $\pi$.
Table~\ref{tab:bounds} compares $d$ with the exponents of the standard worst-case bounds on $\beta$ on four real graphs, where $d$ is $3.7\%$ to $6.4\%$ of $\sqrt{m}$.
Frequently used notations are summarised in Table~\ref{tab:notation}.

\begin{table}[t]
\caption{Notations}
\label{tab:notation}
\centering\footnotesize
\begin{tabular}{ll}
\toprule
Notation & Definition \\
\midrule
$N(u)$, $\Gm(X)$ & neighbours of $u$; common neighbours of $X$ \\
$d$, $\pi$, $\pos(v)$ & degeneracy, degeneracy order, rank of $v$ in $\pi$ \\
$\outv(v)$, $\inv(v)$ & forward and backward neighbours of $v$ \\
$G_{v}$, $\anc{v}$ & residual at $v$; anchored subgraph of $v$ \\
$C$ & candidates of the anchor $v$, i.e., $\outv(v)$ \\
$\Gm_{v}(B)$, $\sat(B)$ & companions adjacent to all of $B$; saturation of $B$ \\
$\beta$, $N$ & numbers of maximal bicliques and of search tree nodes \\
\bottomrule
\end{tabular}
\end{table}

\begin{table}[t]
\caption{Degeneracy and the exponents of the standard bounds on $\beta$}
\label{tab:bounds}
\centering\footnotesize
\begin{tabular}{lrrrrr}
\toprule
graph & $m$ & $\sqrt{m}$ & $\min(|L|,|R|)$ & $d$ & $d/\sqrt{m}$ \\
\midrule
youtube & 293,360 & 542 & 30,087 & 20 & 3.7\% \\
github & 440,237 & 664 & 56,519 & 39 & 5.9\% \\
TV & 3,232,134 & 1,798 & 64,415 & 115 & 6.4\% \\
LJ 108-core & 5,451,039 & 2,335 & 6,662 & 108 & 4.6\% \\
\bottomrule
\end{tabular}
\end{table}

\section{Degeneracy-Anchored Enumeration}\label{sec:algo}

In this section, we propose \degcbo.
We first show that every maximal biclique has a unique anchor and is a local maximal biclique of that anchor.
Then, we characterise what maximality in $G$ requires once the enumeration is confined to the anchored subgraph.
Finally, we give the algorithm, its correctness proof and its bounds.

\subsection{Anchoring}

We first introduce the definitions below.

\begin{definition}[Anchored subgraph]\label{def:anc}
Given an anchor $v$, the \emph{candidates} of $v$ are $C=\outv(v)$, which lie in the vertex set opposite to $v$.
The \emph{companions} of $v$ are the vertices $x$ of $G_{v}$ in the same vertex set as $v$ that are adjacent to at least one candidate.
The \emph{anchored subgraph} of $v$, denoted by $\anc{v}$, is the bipartite subgraph of $G_{v}$ induced by the companions and the candidates.
\end{definition}

Note that $v$ itself is a companion adjacent to every candidate.
For instance, in Figure~\ref{fig:example}, for the anchor $b$, $C=\{1,2\}$ and the companions are $\{a,b,c\}$.

\begin{definition}[Saturation]\label{def:sat}
Given $B\subseteq C$, let $\Gm_{v}(B)=\Gm(B)\cap V(G_{v})$ be the companions adjacent to every vertex of $B$.
The \emph{saturation} of $B$, denoted by $\sat(B)$, is $\Gm(\Gm_{v}(B))\cap C$, i.e., the candidates adjacent to every vertex of $\Gm_{v}(B)$.
$B$ is \emph{saturated} if $\sat(B)=B$.
\end{definition}

Note that $\sat(B)\supseteq B$, and $(\Gm_{v}(B),\sat(B))$ is the biclique of $\anc{v}$ generated by $B$.
If $B$ is saturated, $(\Gm_{v}(B),B)$ is a maximal biclique of $\anc{v}$.
Every maximal biclique of $\anc{v}$ with a non-empty candidate side arises this way, since $v$ is adjacent to all of $C$ and so belongs to every such biclique.
We call the maximal bicliques of $\anc{v}$ the \emph{local} maximal bicliques of $v$, to distinguish them from the maximal bicliques of $G$.
For instance, in Figure~\ref{fig:example}, for the anchor $b$, $\Gm_{b}(\{1\})=\{a,b\}$ and $\sat(\{1\})=\{1,2\}$, i.e., $\{1\}$ is not saturated.
In contrast, $\sat(\{2\})=\{2\}$ and $(\{a,b,c\},\{2\})$ is a local maximal biclique of $b$.

\noindent\textbf{Canonicity test~\cite{kuznetsov1999cbo}.}
The local maximal bicliques are enumerated depth-first by their candidate sides.
From a saturated $B$ and a candidate $c\notin B$, the search forms $B'=\sat(B\cup\{c\})$ and descends into $B'$ if and only if $B'$ and $B$ agree on every candidate smaller than $c$ in a fixed candidate order.
Otherwise, the attempt is a \emph{canonicity failure}, since the biclique with candidate side $B'$ is generated elsewhere.
This is the canonicity test of Close-by-One (\cbo).
It visits every saturated set exactly once, i.e., at the attempt whose candidate $c$ is the smallest candidate of $B'\setminus B$ along its canonical path.
Besides, it does not need a set of vertices that have been expanded previously.

The lemma below confines one side of every maximal biclique to the candidates of a single vertex.

\begin{lemma}[Anchor~\cite{eppstein1994arboricity}]\label{lem:anchor}
Let $(A,B)$ be a maximal biclique of $G$ and let $v$ be the $\pi$-first vertex of $A\cup B$.
If $v\in A$ then $B\subseteq\outv(v)$, and if $v\in B$ then $A\subseteq\outv(v)$.
\end{lemma}
\begin{proof}
W.l.o.g., let $v\in A$. Every $b\in B$ is adjacent to $v$ and $\pos(b)>\pos(v)$, so $b\in\outv(v)$.
\end{proof}

By Lemma~\ref{lem:anchor}, every maximal biclique is assigned to exactly one anchor $v$, i.e., the first vertex of its vertex set, and its side opposite to $v$ is confined to the candidates $C=\outv(v)$, of which there are at most $d$.
Since every vertex of $A\cup B$ is ranked at or after $v$, the side containing $v$ consists of companions of $v$.
As such, $(A,B)$ is a local maximal biclique of $v$, i.e., $B$ is saturated and $A=\Gm_{v}(B)$.
\degcbo\ enumerates, for every anchor $v$, the saturated subsets of $C$ with the \cbo\ canonicity test.
For instance, in Figure~\ref{fig:example}(b), the seven maximal bicliques of $G$ are each reported exactly once, i.e., at the anchor that is the $\pi$-first vertex of their vertex set.
The biclique $(\{a,b\},\{1\})$ is never produced since $\{1\}$ is not saturated at anchor $b$.

\subsection{Maximality beyond the anchored subgraph}

Confining the companions to $G_{v}$ bounds the anchored subgraph.
However, it also breaks maximality.
This is because a local maximal biclique $(X,Y)$ of $v$ is a biclique of $G$, but a vertex ranked before $v$ may extend it.
We call such vertices \emph{ghosts}.
There are two kinds of ghosts, which are characterised by the two lemmas below.

\begin{lemma}[Candidate-side ghosts are exactly $\inv(v)$]\label{lem:colghost}
Let $(X,Y)$ be a local maximal biclique of $v$, so $Y\subseteq C$ and $v\in X$.
Every vertex $w\notin Y$ adjacent to all of $X$ satisfies $w\in\inv(v)$.
\end{lemma}
\begin{proof}
$w$ is adjacent to $v\in X$, so $w\in N(v)=\inv(v)\cup\outv(v)$.
If $w\in\outv(v)=C$ then $w$ is a candidate adjacent to every vertex of $X=\Gm_{v}(Y)$, so $w\in\sat(Y)=Y$, a contradiction.
\end{proof}

\begin{lemma}[Anchor-side ghosts]\label{lem:rowghost}
Let $(X,Y)$ be a local maximal biclique of $v$.
Every vertex $u\notin X$ adjacent to all of $Y$ satisfies $\pos(u)<\pos(v)$.
\end{lemma}
\begin{proof}
A vertex adjacent to all of $Y$ with $\pos(u)\ge\pos(v)$ is a companion in $\Gm_{v}(Y)=X$.
\end{proof}

For instance, in Figure~\ref{fig:example}(b), at anchor $1$, the candidate-side ghost $b\in\inv(1)$ does not kill $\{a\}$ because $b$ is not adjacent to the companion $3$.
In contrast, at anchor $c$, the local maximal biclique $(\{c\},\{3\})$ is extended by the candidate-side ghosts $2,4\in\inv(c)$ and by the anchor-side ghosts $a$ and $e$.
The two lemmas lead to the reporting rule below.

\begin{definition}[Reporting rule]\label{def:emit}
A local maximal biclique $(X,Y)$ of $v$ is \emph{canonical at $v$} if no $w\in\inv(v)$ is adjacent to all of $X$ and no $u$ with $\pos(u)<\pos(v)$ is adjacent to all of $Y$.
\degcbo\ reports a local maximal biclique if and only if it is canonical.
\end{definition}

Both tests are necessary.
Without the candidate-side ghost test, bicliques extendable by a backward neighbour of $v$ are reported.
Without the anchor-side ghost test, bicliques extendable by an earlier vertex in the same vertex set as $v$ are reported.
For instance, on \textsf{youtube}, the second omission reports $3{,}417{,}456$ bicliques instead of the correct $1{,}826{,}587$.
Both tests are also cheap.
Since the candidate-side ghosts of $v$ are the fixed set $\inv(v)$, for every companion $x$ we precompute a bitmask over $\inv(v)$ of the ghosts adjacent to $x$, and the test for $X$ is the aggregate AND of the masks of its companions with early exit.
The anchor-side ghosts of $v$ are the earlier vertices in the same vertex set as $v$ adjacent to some candidate.
For every such vertex, we precompute a bitmask over $C$, and the test for $Y$ asks whether some mask contains $Y$.
Section~\ref{sec:impl} shows that the anchor-side ghost masks can be reduced to their maximal elements without loss, which reduces their number by one to two orders of magnitude.

\subsection{The algorithm}

\begin{algorithm}[t]
\small
\Input{bipartite graph $G=(L,R,E)$}
\Output{every maximal biclique of $G$, exactly once}
compute a degeneracy order $\pi$ and rename vertices by rank\;
\ForEach{$v$ in increasing $\pos(v)$}{
  $C\leftarrow\outv(v)$ sorted by non-decreasing degree\;
  \Setup{$v$, $C$}: companions $X_{0}$ with candidate masks $\mathit{adj}[x]\subseteq C$, ghost masks $\mathit{gh}[x]\subseteq\inv(v)$, maximal anchor-side ghost masks $\mathcal{R}$\;
  \Enum{$\emptyset$, $X_{0}$, $0$, $\mathcal{R}$}\;
}
\BlankLine
\SetKwProg{Fn}{Procedure}{}{}
\Fn{\Enum{$Y$, $X$, $c_{0}$, $\mathcal{R}$}}{
  \ForEach{candidate $c\ge c_{0}$ with $c\notin Y$, in decreasing order}{
    \lIf{$\mathit{NF}[c]\ne\bot$ and $\mathit{NF}[c]$ has a candidate $<c$ outside $Y$}{\textbf{continue}}
    $X'\leftarrow\{x\in X: c\in\mathit{adj}[x]\}$; \quad $Y'\leftarrow\bigwedge_{x\in X'}\mathit{adj}[x]$ \tcc*{one pass}
    \lIf{$Y'\setminus Y$ has a candidate $<c$}{$\mathit{NF}[c]\leftarrow Y'$; \textbf{continue}}
    \lIf{$\bigwedge_{x\in X'}\mathit{gh}[x]\ne 0$}{\textbf{continue} \tcc*[f]{candidate-side ghost: prune subtree}}
    $\mathcal{R}'\leftarrow\{M\in\mathcal{R}: Y'\subseteq M\}$\;
    \lIf{$\mathcal{R}'=\emptyset$}{report $(X'\cup\{v\}, Y')$}
    \lElseIf{every $x\in X'\setminus\{v\}$ has $\mathit{adj}[x]\subseteq M$ for some $M\in\mathcal{R}'$}{\textbf{continue} \tcc*[f]{barren}}
    \Enum{$Y'$, $X'$, $c+1$, $\mathcal{R}'$}\;
  }
  restore the entries of $\mathit{NF}$ written in this call\;
}
\caption{\degcbo$(G)$}\label{alg:degcbo}
\end{algorithm}

\noindent\textbf{The algorithm.}
Algorithm~\ref{alg:degcbo} gives \degcbo\ at the level of bitmasks, which is the level at which it is implemented.
After the order is computed, vertices are renamed by rank so that $\pos$ is the identity and every adjacency list is sorted.
As such, the backward and forward neighbours of $v$ are the prefix and suffix of its list around $v$, found by one binary search.
For every anchor, the set-up (line 4, Section~\ref{sec:setup}) builds the anchored subgraph, i.e., the candidate masks of the companions over $C$, the ghost masks over $\inv(v)$, and the maximal anchor-side ghost masks.
The recursion (lines 7 to 16) extends the candidate side one vertex at a time and saturates.
Besides the \cbo\ canonicity test and the two ghost tests, it includes three refinements discussed in Sections~\ref{sec:setup} and \ref{sec:impl}.
First, candidates are visited in decreasing order and failed saturations are remembered in $\mathit{NF}$ (lines 8 and 10), i.e., the failure propagation of Section~\ref{sec:setup}.
Second, the companion filter and the saturation are computed in one pass over the companions (line 9), i.e., the mask transpose.
Third, a node whose every companion lies inside a surviving anchor-side ghost mask is barren and its subtree is skipped (line 14).
Here, a \emph{node} of the search tree corresponds to a recursive subproblem in the terminology of~\cite{chen2022oombea}.

\noindent\textbf{Correctness.}
We are ready to show that Algorithm~\ref{alg:degcbo} is correct.

\begin{lemma}[Correctness]\label{thm:correct}
Algorithm~\ref{alg:degcbo} reports every maximal biclique of $G$ exactly once.
\end{lemma}
\begin{proof}
\emph{Completeness.} Let $(A,B)$ be maximal with anchor $v$, and w.l.o.g.\ $v\in A$.
By Lemma~\ref{lem:anchor} $B\subseteq C$; every $a\in A$ is ranked at or after $v$, so $A\subseteq G_{v}$; and $A=\Gm(B)$ by maximality, so $A=\Gm_{v}(B)$ and $B$ is saturated.
The search visits every saturated subset of $C$ exactly once, so it visits $B$ with companion side $A$.
No candidate-side ghost or anchor-side ghost extends $(A,B)$, since it is maximal, so the node passes both tests and is reported.
Barren pruning never skips a reportable node: if every companion of $X'$ lies inside some surviving anchor-side ghost mask, then every descendant's candidate side $Y''\subseteq\mathit{adj}[x]$ for some companion $x$ of its side lies inside that mask, so every descendant fails the anchor-side ghost test.
Failure propagation skips an attempt only when its saturation is guaranteed to fail canonicity (Lemma~\ref{lem:fcbo}).
\emph{Uniqueness.} The anchor of $(A,B)$ is unique, and the residual confines the vertex set to anchors ranked at or before every vertex of $A\cup B$, so $(A,B)$ can only be produced at $v$; within $v$, canonicity produces each saturated set once.
\end{proof}

\subsection{Bounds}

\noindent\textbf{Time complexity.}
We first bound the size of the search tree.

\begin{theorem}[Search space]\label{thm:space}
Let $N$ be the number of nodes of the search tree over all anchors. Then
$\beta\le N\le\sum_{v}2^{|\outv(v)|}\le n\cdot 2^{d}$.
\end{theorem}
\begin{proof}[Proof sketch]
Every node is a distinct saturated subset of $C_{v}$, of which there are at most $2^{|C_{v}|}\le 2^{d}$, and every reported biclique occupies its own node.
\end{proof}

The lemma below shows that the search tree is also bounded by the output.

\begin{lemma}[Output sensitivity]\label{thm:output}
$N\le\sum_{v}\beta(\anc{v})\le n\cdot\beta$, so Algorithm~\ref{alg:degcbo} runs in time polynomial in $n$ and $\beta$, with or without its prunings.
\end{lemma}
\begin{proof}
Every node at anchor $v$ is a local maximal biclique of $\anc{v}$, which is an induced subgraph of $G$.
A maximal biclique $(A,B)$ of an induced subgraph $H$ extends to a maximal biclique $(A',B')$ of $G$, and $(A'\cap V(H),B'\cap V(H))$ is a biclique of $H$ containing $(A,B)$, hence equal to it; so extension is injective and $\beta(H)\le\beta(G)$.
Each node costs polynomial time and makes at most $|C_{v}|\le d$ failed attempts.
\end{proof}

\noindent\textbf{Discussion.}
Note that the two bounds are incomparable, i.e., $n\cdot2^{d}$ is small when the graph is sparse, whereas $n\cdot\beta$ is small when the output is small.
As such, the search tree is bounded by the smaller of the two.
For instance, on the real graphs, the measured $N/\beta$ is far below $n$, i.e., between $1.0$ and $2.8$.
The lemma below bounds the degeneracy of a bipartite graph.

\begin{lemma}[Bipartite degeneracy]\label{thm:deg}
In a bipartite graph $d\le\min\{\sqrt{m},\min(|L|,|R|)\}\le\sqrt{|L||R|}\le(|L|+|R|)/2$, and the first two inequalities are tight.
\end{lemma}
\begin{proof}
Let $H=(L',R',E')$ be a subgraph of minimum degree $d$.
Every $u\in L'$ has $d$ neighbours in $R'$, so $|R'|\ge d$, and symmetrically $|L'|\ge d$; hence $d\le\min(|L|,|R|)$.
Also $m\ge|E'|\ge|L'|\cdot d\ge d^{2}$.
The remaining inequalities are $m\le|L||R|$ and the arithmetic--geometric mean inequality.
The $k$-crown $K_{k,k}$ minus a perfect matching has $d=k-1$ and $m=k^{2}-k$; $K_{k,k}$ has $d=k=\min(|L|,|R|)$.
\end{proof}

The generic bound for arbitrary graphs is $d\le\sqrt{2m}$.
The bipartite strengthening matters, since with it $n\cdot2^{d}$ is dominated by none of the three classical bounds $2^{\sqrt m}$~\cite{schutt1987}, $2^{\min(|L|,|R|)}$ and $2^{(|L|+|R|)/2}$, and in fact implies the first as a corollary.

\begin{corollary}[Worst case]\label{cor:opt}
\degcbo\ computes $\beta$ in $\OO^{*}(2^{\sqrt{m}})$ time.
For enumeration this is optimal, since the crowns have $\beta=2^{\Theta(\sqrt{m})}$ and must be written; on the $k$-crown $N=\beta+k-2$ exactly.
\end{corollary}

\noindent\textbf{Remark.}
For counting, the bound is not tight.
This is because the crowns are counted in closed form, and Section~\ref{sec:count} shows that the search tree of the $k$-crown shrinks to $\OO(k^{2})$ nodes once star-shaped subtrees are counted by formula.
Besides, a worst-case bound alone is weak, since an algorithm that spends $2^{\sqrt m}$ on every input satisfies it.
Lemma~\ref{thm:output} is the complementary statement.
For instance, on $K_{120,120}$, where $\beta=1$ and the exponential bound allows $2^{120}$, the search tree has $239$ nodes.
Finally, we bound the cost of a node.

\begin{lemma}[Per-node cost]\label{thm:node}
A node with companion side $X$ costs $\OO(|X|+g_{v}+\lceil|\inv(v)|/64\rceil)$ word operations, where $g_{v}$ is the number of surviving anchor-side ghost masks.
\end{lemma}
\begin{proof}[Proof sketch]
One pass over the parent's companions computes the filtered companion side and its saturation as a running AND of candidate masks (line 9); canonicity is one mask operation; the candidate-side ghost test is an aggregate AND over $\lceil|\inv(v)|/64\rceil$ words per companion with early exit; the anchor-side ghost filter touches the surviving list, which shrinks as $Y$ grows.
\end{proof}

As such, the saturation takes $\OO(|X|)$ word operations instead of $\OO(|C|\cdot|X|)$ set intersections.
This is because a candidate belongs to the saturation if and only if its bit survives the AND of the companion masks.
We call this the \emph{mask transpose}.
It answers the saturation, the canonicity witness and the extender test with a single pass over the companions.
However, the bounds do not account for the cost of building the anchored subgraph at every anchor, which is the subject of the next section.

\section{The Set-Up Problem and the Cost-Aware Sweep}\label{sec:setup}

In this section, we study the set-up problem, i.e., the cost of building the anchored subgraph at every anchor, and solve it on the anchored side of the design with three exact techniques.

\noindent\textbf{The set-up scan.}
Building the anchored subgraph of $v$ by scanning the adjacency list of every candidate costs $\sum_{c\in\outv(v)}\deg(c)$.
Summed over all anchors, every vertex $c$ is charged once per backward neighbour, i.e.,
\begin{equation}\label{eq:scan}
\sum_{v}\sum_{c\in\outv(v)}\deg(c)\;=\;\sum_{c}\deg(c)\cdot|\inv(c)|\;=\;\OO(\Delta m).
\end{equation}
The degeneracy orientation bounds $|\outv(c)|$ but not $|\inv(c)|$.
As such, this term behaves like $\sum_{c}\deg(c)^{2}$ on hub-heavy graphs.
It is not an artefact of our implementation.
On \textsf{wikicat}, a graph of $1.85$ million articles and $183$ thousand categories, the degeneracy order removes the low-degree articles first.
As such, almost every anchor is an article whose candidates are hub categories, and the scan touches $1.5$ billion adjacency entries for a search tree of $7$ million nodes.
This scan was $86.8\%$ of the running time, and the enumeration itself was $6.7\%$.
Sweeping in the other direction, i.e., from the hub categories through the low-degree articles, costs about $2m$.
However, it is not available inside the bitmask design.
This is because anchoring on a hub makes its co-occurring categories the candidates, thousands of them, and loses the bounded anchored subgraph that gives Theorem~\ref{thm:space}.

\subsection{Pair-list set-up}

\noindent\textbf{Intuition.}
Let $k(x)=|N(x)\cap C|$ be the number of the anchor's candidates adjacent to a companion $x$.
We first introduce the following observation.
\begin{observation}\label{obs:single}
A companion $x$ with $k(x)=1$, adjacent to the single candidate $c$, belongs to exactly one local maximal biclique, i.e., the one with candidate side $\{c\}$, and to none with two or more candidates.
The companions with $k(x)\ge 2$ are exactly the common neighbours of some pair of candidates.
\end{observation}
On \textsf{wikicat}, the companions with $k(x)=1$ are $87\%$ of all companions and, since they are the neighbours of hub candidates, nearly all of the scan.

\noindent\textbf{Pair lists.}
For two vertices $c_{1},c_{2}$ on the same side, their common neighbours sorted by rank are denoted by $\comm(c_{1},c_{2})=N(c_{1})\cap N(c_{2})$.
We precompute the lists $\comm(c_{1},c_{2})$ once for every pair with at least one common neighbour.
To do so, every vertex $x$ of the other side emits the $\binom{\deg(x)}{2}$ pairs of its neighbours, and the triples are sorted.
The total size is $\sum_{x}\binom{\deg(x)}{2}$ over that side, which is $4.7$ million entries on \textsf{wikicat}.
The lists are indexed by a sorted partner array per vertex, so a list is found by one binary search.
We choose the side whose vertices give the smaller sum.
Whether to build the lists at all is decided by sampling.
Specifically, for a few thousand anchors, we compute the scan cost and the pair cost exactly from the adjacency lists and scale the saving to all anchors.
The lists are built only if the predicted saving exceeds the build cost, which is proportional to their size.
Otherwise, the algorithm scans.
The lists are built on \textsf{wikicat}, \textsf{bibsonomy} and \textsf{twitter}, but not on \textsf{youtube}, \textsf{dblp} and \textsf{bookcrossing}, where the companions have high degree and the build would cost more than it saves.

\noindent\textbf{The set-up.}
With pair lists, the anchored subgraph of $v$ is built as follows.
For every pair $c_{i},c_{j}$ of candidates, we walk $\comm(c_{i},c_{j})$, which one binary search splits at $v$.
Entries ranked before $v$ become anchor-side ghosts with bits $i$ and $j$, and entries ranked after $v$ become companions with bits $i$ and $j$.
A stamp array merges the contributions of different pairs to the same vertex.
This materialises exactly the companions and anchor-side ghosts with two or more bits.
For instance, for anchor $b$ in Figure~\ref{fig:example}, the only pair list walked is $\comm(1,2)=\{a,b\}$, which materialises the companion $a$ with both bits.
In contrast, the companion $c$ is adjacent to candidate $2$ alone, i.e., $k(c)=1$, and it is never materialised.
We account for the companions with one bit by counting.
Let $\suf(c)$ be the number of vertices ranked after $v$ in $N(c)$ and let $r_{i}$ be the number of materialised companions with bit $i$.
Then, $n_{1}(c_{i})=\suf(c_{i})-1-r_{i}$ companions are adjacent to $c_{i}$ alone, and similarly $g_{1}(c_{i})$ earlier vertices are adjacent to $c_{i}$ alone.
We are ready to show the lemma below.
\begin{lemma}[Single-candidate companions]\label{lem:single}
Let the set-up materialise only companions and anchor-side ghosts with two or more bits, and let the recursion be modified at the root only as follows:
1) if $n_{1}(c)>0$, the saturation of $\{c\}$ is $\{c\}$;
2) if $g_{1}(c)>0$, an anchor-side ghost mask $\{c\}$ is added; and
3) $w\in\inv(v)$ kills $\{c\}$ if and only if $|\comm(c,w)\cap\{x:\pos(x)>\pos(v)\}|=\suf(c)-1$.
Then Algorithm~\ref{alg:degcbo} visits the same nodes, with the same saturations, canonicity outcomes, prunes and reports, as with the full anchored subgraph.
\end{lemma}
\begin{proof}
A single-candidate companion $x$ adjacent to $c$ lies in $\Gm_{v}(Y)$ for a candidate set $Y$ only if $Y\subseteq\{c\}$, so it affects exactly the root child $\{c\}$.
Its effect on the saturation of $\{c\}$ is to veto every other candidate, which the forced saturation reproduces; if $n_{1}(c)=0$ the saturation over the materialised companions is the true saturation.
An earlier vertex adjacent to $c$ alone covers exactly the candidate sets $Y\subseteq\{c\}$, which the mask $\{c\}$ reproduces, and it participates in the containment reduction and the barren test exactly as a materialised ghost would.
A candidate-side ghost $w$ kills $\{c\}$ if and only if it is adjacent to every neighbour of $c$ ranked after $v$ and to $v$; $w\in\inv(v)$ is adjacent to $v$, and the count condition states the rest.
The reachable-candidate set and the barren test at the root child use only companions with a candidate outside $\{c\}$, which are materialised.
Deeper nodes have companion sides of vertices with two or more bits.
\end{proof}

\noindent\textbf{Ghost masks.}
The ghost masks for the materialised companions are obtained from the same structure.
Specifically, the neighbours of $c_{i}$ adjacent to $w\in\inv(v)$ are $\comm(c_{i},w)$ above $v$, so one walk per $(i,w)$ pair sets the bits, instead of a scan of $N(w)$ for every $w$.

\noindent\textbf{Cost-aware choice.}
Per anchor, the scan costs $\sum_{c}\deg(c)+\sum_{w\in\inv(v)}\suf(w)$ touches and the pair set-up costs $\sum_{i<j}|\comm(c_{i},c_{j})|+\sum_{i,w}|\comm(c_{i},w)|$ touches plus one binary search per pair.
Both are known before any work is done, since the list lengths are stored, and \degcbo\ chooses the cheaper.
A lookup is charged as eight touches, and the pair test is skipped altogether when the scan is small or the lookups alone would exceed it.
On \textsf{wikicat}, the pair set-up is chosen for $451$ thousand anchors, walks $222$ million list entries, saves $634$ million scan touches and leaves $278$ million single-candidate companions implicit.
The pair set-up realises, inside the anchored design, the cost-aware sweep direction, which was the weakness of the method.
This is because the scan sweeps candidates and pays for every neighbour.
In contrast, the pair walk sweeps the companions that matter and pays $\binom{k(x)}{2}$ per materialised companion, which is a constant on sparse sides.

\noindent\textbf{Time complexity.}
The sum of list lengths over all anchors is $\sum_{x}\sum_{v}[\pos(v)<\text{all of } N(x)\cap C_{v}]\binom{k(x)}{2}$, which is bounded by $\sum_{x}\binom{\deg(x)}{2}\cdot\max|\inv|$.
In practice, it is $4.6$ to $11.5$ times smaller than Equation~\eqref{eq:scan} on the graphs where the scan dominates.
Next, we reduce the input itself before any anchor is processed.

\subsection{Twin quotient}

\begin{definition}[Twins]\label{def:twin}
Two vertices on the same side are \emph{twins} if they have the same neighbourhood.
\end{definition}
The lemma below shows that merging twins is exact.
\begin{lemma}[Twin quotient]\label{lem:twin}
Let $G^{*}$ be obtained from $G$ by merging every twin class into one vertex.
The maximal bicliques of $G$ and of $G^{*}$ are in bijection, so $\beta(G)=\beta(G^{*})$.
\end{lemma}
\begin{proof}
A maximal biclique $(A,B)$ with $u\in A$ contains every twin $u'$ of $u$, since $N(u')=N(u)\supseteq B$ and $(A\cup\{u'\},B)$ is a biclique; so $A$ and $B$ are unions of twin classes, and merging classes maps $(A,B)$ to a biclique of $G^{*}$ that is maximal for the same reason.
Conversely a maximal biclique of $G^{*}$ expands to a maximal biclique of $G$.
Twin classes on one side are unaffected by merging the other side, since a class is adjacent to a vertex either entirely or not at all, so one pass per side suffices.
\end{proof}

\noindent\textbf{Time complexity.}
The quotient is computed in $\OO(m\log n)$ by hashing sorted adjacency lists and verifying collisions.
Since it is a reduction that every \mbe\ algorithm may apply, we report \ips\ both on the original graphs and on the quotients.
On \textsf{wikicat}, it removes $60\%$ of the articles and $38\%$ of the edges and halves the sum of squared hub degrees.
On \textsf{bibsonomy}, it removes $58\%$ of the items.
Counting needs no multiplicities, so nothing is kept beyond the quotient.
Next, we turn to the saturations that fail canonicity in the search.

\subsection{Effective failure propagation}

The search computes a saturation for every attempt.
However, in the hub bands of real graphs most attempts fail canonicity, i.e., there are $2.2$ failures per reported biclique on average over the hub anchors of \textsf{flickr}, $6.4$ on the hardest anchor, and $7.0$ on \textsf{TV}.

\noindent\textbf{The rule of \fcbo.}
\fcbo~\cite{outrata2012fcbo} remembers, for every candidate $c$, the saturation $\mathit{NF}[c]$ that made $c$ fail at an ancestor.
Then, the attempt at a descendant is skipped when $\mathit{NF}[c]$ still contains a candidate smaller than $c$ outside the current set $Y$.
This is because the new saturation contains the old one, so it fails again, as shown in the lemma below.
\begin{lemma}[Failure propagation]\label{lem:fcbo}
If $\sat(Y_{0}\cup\{c\})$ contains a candidate $i<c$ with $i\notin Y_{0}$, and $Y\supseteq Y_{0}$ with $i\notin Y$, then $\sat(Y\cup\{c\})\ni i$ and the attempt at $c$ from $Y$ fails canonicity.
\end{lemma}
\begin{proof}
Saturation is monotone: $Y\cup\{c\}\supseteq Y_{0}\cup\{c\}$ implies $\sat(Y\cup\{c\})\supseteq\sat(Y_{0}\cup\{c\})\ni i$.
\end{proof}

\noindent\textbf{Shortcoming.}
We implemented this rule and measured zero skips on every graph.
Consider a depth-first search that recurses into each child inside the loop iteration that creates it and visits candidates in increasing order.
A failure at candidate $c$ is then recorded only after every child created at candidates $c'<c$ has already been explored.
Note that these children are the only descendants that ever re-attempt $c$.
As such, the propagation is structurally vacuous.

\noindent\textbf{Decreasing order.}
Visiting the candidates of a node in \emph{decreasing} order (line 7) records the failure at $c$ before the children at $c'<c$ are created, with the same tree, the same bicliques and the same number of nodes.
With this one change, $62$ to $99\%$ of the failed saturations are skipped where failures are frequent.
Besides, the failure array is restored on return (line 16) so that the rule stays sound.
In the next section, we show that counting need not visit one node per biclique.

\section{Counting Beyond Enumeration}\label{sec:count}

The algorithm of Section~\ref{sec:algo} counts by visiting, i.e., every reported biclique is a node of the search tree, so $N\ge\beta$.
On the crowns, where $\beta=2^{k}-2$, the tree has $2^{k}+k-4$ nodes.
For enumeration this is unavoidable, since the output must be written.
However, for counting it is not.
This is because the crown count is a closed form, and nothing in Theorem~\ref{thm:space} forces a counter to visit $2^{\Theta(\sqrt m)}$ nodes.
It is natural to ask whether a counter can avoid visiting a node per biclique.
In this section, we give two exact answers.
Firstly, we collapse a class of subtrees to a formula, which turns the crowns from the worst case into a trivial one and, on the real graphs, measures how much of $\beta$ lies in subtrees with that structure.
Secondly, we show that the recent sub-exponential algorithms for counting satisfying assignments give a counting bound that no enumerator can match, and we state this result for the bound it implies.
Finally, we prove a conditional lower bound for counting in $d$.

\subsection{Closed-form subtrees}\label{sec:cf}

\noindent\textbf{Overall idea.}
Consider a node of the search at anchor $v$ with candidate side $Y$, companion side $X\ni v$ and branch point $c$, so that its subtree can only add candidates above $c$.
A candidate is \emph{free} if it is above $c$, outside $Y$ and present in some companion other than $v$, and the set of free candidates is denoted by $F$.
A candidate is \emph{low} if it is below $c$ and outside $Y$.
A companion other than $v$ is \emph{safe} if it contains every free candidate.
The subtree enumerates the saturated sets $Z\supsetneq Y$ that agree with $Y$ below $c$.
We are ready to show when their number is a formula.

\begin{theorem}[Star-shaped subtrees]\label{thm:star}
Let the node have no surviving anchor-side ghost and let every companion other than $v$ miss at most one free candidate. Then the subtree reports exactly $2^{|F|}-1$ bicliques, plus one, in either of the following cases:
\begin{enumerate}
\item[(a)] some companion is safe, every low candidate is missed by a safe companion, and no backward neighbour of $v$ is adjacent to every safe companion. The extra biclique is $(\{v\},C)$, and it is present if and only if $v$ has no backward neighbour, there is no low candidate, and some candidate above $c$ outside $Y$ lies in no companion other than $v$;
\item[(b)] no companion is safe, $v$ has no backward neighbour, and $Y\cup F=C$; there is no extra biclique.
\end{enumerate}
\end{theorem}
\begin{proof}
A descendant with a companion $x\ne v$ on its side has candidate side within $\mathit{adj}[x]$, and as such within $Y\cup F$ once the low candidates are excluded by canonicity; a descendant whose side is $\{v\}$ alone has candidate side $C$.
For $S\subseteq F$ the companions adjacent to all of $Y\cup S$ are $v$, the safe companions, and the companions whose single missed candidate lies outside $S$.
Each free candidate $f\notin S$ is missed by some companion, since $f\notin Y$ and $Y$ is saturated, and that companion misses only $f$, so it survives $S$ and keeps $f$ out of the saturation; a surviving safe companion keeps every low candidate it misses and every candidate above $c$ absent from all companions out of it.
Under (a) a safe companion always survives, so $\sat(Y\cup S)=Y\cup S$ for every non-empty $S$, the $2^{|F|}-1$ sets are distinct, they agree with $Y$ below $c$, and none is extended by a backward neighbour of $v$, since such a neighbour would have to be adjacent to every safe companion, nor by an anchor-side ghost, since none survives.
The set $C$ is generated exactly once on the canonical path through the candidates above $c$ in increasing order, i.e., when that path first meets a candidate absent from every companion other than $v$.
It is canonical if and only if there is no low candidate.
It is reported if and only if no backward neighbour of $v$ exists, since every backward neighbour is adjacent to $v$.
Under (b) a companion missing some $f\notin S$ survives whenever $S\ne F$, and for $S=F$ the side is $\{v\}$ and the saturation is $C=Y\cup F$; so again every non-empty $S$ gives a distinct saturated set, and no further set exists.
\end{proof}

\noindent\textbf{Time complexity.}
The test costs one pass over the companions, which is the same pass that the node performs anyway, and the subtree is never entered when the test fires.
For instance, on the $k$-crown every companion misses exactly its matched vertex.
As such, the anchored subgraph of every anchor is star-shaped and the terminal fires at every child of the root, which leads to the following corollary.
\begin{corollary}[Crowns]\label{cor:crown}
With the terminal, the search tree of the $k$-crown has $\OO(k^{2})$ nodes.
\end{corollary}
On the $20$- and $24$-crowns, the tree has $155$ and $222$ nodes for $1{,}048{,}574$ and $16{,}777{,}214$ bicliques, against $2^{20}+16$ and $2^{24}+20$ without the terminal.
As such, the family that forces $2^{\sqrt m}$ on every enumeration algorithm is counted in polynomial time.
We evaluate the terminal on the real graphs in Section~\ref{sec:exp}.

\noindent\textbf{Discussion.}
The terminal is one instance of a general principle, i.e., a subtree whose companions miss the free candidates in a pattern of bounded complexity has a count that is a function of the pattern.
The star pattern is the one that real graphs exhibit.
A richer pattern requires a richer formula, and the output-bound instances exhibit none.
Next, we study what can be said about counting in general.

\subsection{Counting in sub-exponential time}\label{sec:subexp}

Two October 2026 preprints claim deterministic $2^{o(n)}$-time algorithms that count the satisfying assignments of any $k$-CNF on $n$ variables, refuting the Exponential Time Hypothesis (ETH)~\cite{openai2026eth,anthropic2026sat,kintali2026eth}.
Note that they carry machine-checked formalisations but have not yet been independently refereed.
As such, the corollary below is conditional on them.
Maximality has clauses of unbounded width, so $\beta$ is not a $k$-CNF count as it stands.
We make it one by inclusion--exclusion over the vertices that would extend a biclique, as shown in the lemma below.

\begin{lemma}[Pairwise form of $\beta$]\label{lem:ie}
Let $\beta'$ count the pairs $(A,B)$ with $A=\Gm(B)$ and $B=\Gm(A)$, including the at most two pairs with an empty side. Then
\[
\beta'=\sum_{A,U}\sum_{B,V}(-1)^{|U|+|V|}\,\bigl[A\times B,\ U\times B,\ A\times V\subseteq E\bigr],
\]
where the sums run over disjoint $A,U\subseteq L$ and disjoint $B,V\subseteq R$.
\end{lemma}
\begin{proof}
A biclique $(A,B)$ is counted by $\beta'$ if and only if no $u\notin A$ is adjacent to all of $B$ and no $v\notin B$ to all of $A$, i.e., $\prod_{u\notin A}(1-[\{u\}\times B\subseteq E])\prod_{v\notin B}(1-[A\times\{v\}\subseteq E])=1$; expanding both products gives the sum.
\end{proof}

The summand is a product of conditions on pairs, i.e., a non-edge $(x,y)$ forbids the state pairs $(A,B)$, $(U,B)$ and $(A,V)$, and an edge forbids nothing.
With two bits per vertex, a constant number of clauses of width four per non-edge and a parity chain for the sign, $\beta'$ is the difference of two $4$-CNF counts on $\OO(n)$ variables.
As such, Theorem~1.1 of~\cite{openai2026eth} applies with arbitrary clause count, which leads to the following corollary.
\begin{corollary}[Sub-exponential counting]\label{cor:subexp}
The number of maximal bicliques of a bipartite graph on $n$ vertices can be computed in $2^{\OO(n\log\log n/\sqrt{\log n})}$ time.
\end{corollary}

\noindent\textbf{Discussion.}
This bound is below the $2^{n/2}$ that the crowns force on every enumeration algorithm, and below the $1.3954^{n}$ branching bound of the state-of-the-art algorithm \ips.
However, Corollary~\ref{cor:subexp} is a bound and not a practical algorithm.
Firstly, the exponent is in $n$, so it improves on $n\cdot 2^{d}$ only when $m\gtrsim n^{2}/\log n$, i.e., the dense regime where $d$ is of order $n$.
Secondly, the underlying algorithms have no practical crossover point, with a structural lower bound of about $2^{6000}$ variables for their whole class~\cite{majdoddin2026crossover}.
Besides, the reduction does not localise.
This is because inside an anchored subgraph the companions become constraints of width up to $d$ over the $d$ candidates, and a model counter whose exponent is in the number of variables must receive the companions as variables too.
As such, its exponent is in the size of the anchored subgraph and not in $d$.

\subsection{Hardness of counting in the degeneracy}\label{sec:hard}

In this subsection, we prove a conditional lower bound for counting in $d$, since counting is not output-bound and the crowns give no lower bound at all.
We prove the bound under the Satisfiability Hypothesis (SH)~\cite{stearnshunt1990}, i.e., CNF-SAT on formulas of length $L$ has no $2^{L^{1-\varepsilon}}$-time algorithm.
SH survived the refutation of ETH and is equivalent to the Weak ETH that replaced it~\cite{mit2026sh}.
We first introduce the following lemma.

\begin{lemma}[Union form of $\beta$]\label{lem:union}
Let $C$ be one side of $G$ and, for every vertex $x$ of the other side, let $E_{x}=C\setminus N(x)$ be its miss set.
A set $Y\subseteq C$ is saturated if and only if $C\setminus Y$ is a union of miss sets.
As such, $\beta$ equals the number of distinct unions of subfamilies of $\{E_{x}\}$, minus one for the empty union when no miss set is empty, minus one more if the miss sets cover $C$.
Besides, every hypergraph on $C$ arises as a family of miss sets.
\end{lemma}
\begin{proof}
$Y$ is saturated if and only if every $c\notin Y$ has a vertex $x$ with $Y\subseteq N(x)$ and $c\notin N(x)$, i.e., $c\in E_{x}\subseteq C\setminus Y$; this says exactly that $C\setminus Y$ is the union of the miss sets it contains.
The two corrections remove $Y=C$ when its companion side is empty and $Y=\emptyset$.
For the converse, give the hypergraph's edges as the other side, each adjacent to the complement of its edge.
\end{proof}

\noindent\textbf{Intuition.}
Lemma~\ref{lem:union} turns counting into a problem about a hypergraph on $s=\min(|L|,|R|)$ vertices whose solution family is closed under union.
This closure blocks a direct encoding of satisfiability, and the reduction in the theorem below circumvents it.

\begin{theorem}[Lower bound]\label{thm:lb}
Assuming SH, for every $\varepsilon>0$ no algorithm computes $\beta$ for every bipartite graph in $2^{s^{1-\varepsilon}}\cdot\mathrm{poly}$ time, where $s=\min(|L|,|R|)$, nor in $2^{d^{1-\varepsilon}}\cdot\mathrm{poly}$ time.
\end{theorem}
\begin{proof}[Proof sketch]
By Lemma~\ref{lem:union} it suffices to count the unions of subfamilies of a hypergraph on $s$ vertices.
Given a CNF $\varphi$ with $n$ variables and $m$ clauses, we add the $n$ clauses $x_{i}\vee\neg x_{i}$.
For a set $S$ of distinct primes, we build the hypergraph on vertices $t_{i},f_{i}$ (literals), $c_{j}$ (clauses) and, for each clause and each $p\in S$, a chain $d^{p}_{j,1},\dots,d^{p}_{j,p-1}$.
Its edges are $\{t_{i},f_{i},u\}$ for every vertex $u$, $\{\ell\}$ for every literal, and $\{c_{j},\ell\}\cup\{d^{p}_{j,1},\dots,d^{p}_{j,q}\}$ for every literal $\ell$ of clause $j$, every $p\in S$ and every prefix length $0\le q\le p-1$.
Every set containing a complementary pair is a union of edges, and there are $2^{|V|}-3^{n}2^{|V|-2n}$ of them.
A set without one is a union of edges if and only if its literals $T$ are arbitrary, its clause vertices are hit by $T$, and its chain vertices are prefixes below present clause vertices, which gives $1+\prod_{p\in S}p$ configurations per hit clause.
As such, the number of unions equals the constant plus $\sum_{k}a_{k}(1+\prod_{p\in S}p)^{k}$, where $a_{k}$ counts the consistent partial assignments hitting exactly $k$ clauses.
Distinct $S$ give distinct products, so $m+n+1$ subsets of the first $\lceil\log_{2}(m+n+2)\rceil$ primes determine the polynomial, whose leading coefficient is the number of total assignments satisfying $\varphi$.
The instances have $s=|V|=(n+m)^{1+o(1)}$, are dense, so $d\ge s-O(\log s\log\log s)$, and are built in polynomial time; a $2^{s^{1-\varepsilon}}$-time counter would therefore decide CNF-SAT in $2^{L^{1-\varepsilon'}}$ time.
\end{proof}

\noindent\textbf{Remark.}
The base of the exponent in $n\cdot2^{d}$ cannot be improved either under a stronger hypothesis.
This is because the reduction of~\cite{kuznetsov2001}, applied to a monotone CNF of any width, has the number of variables as smaller side.
As such, by the hardness of the parity of the number of hitting sets in the size of the universe~\cite{cygan2016}, a $2^{(1-\varepsilon)s}$-time counter would refute $\oplus$SETH.
Note that none of this concerns the sparse inputs of Section~\ref{sec:exp}, on which $d\ll s$ and the search is output-bound.
Besides, the experiments of Section~\ref{sec:exp} show that the bicliques of the hard instances lie in anchored subgraphs with no structure of either kind.
As such, for them counting costs what enumeration costs.

\section{Implementation}\label{sec:impl}

In this section, we present the implementation techniques that realise the bounds of Section~\ref{sec:algo}.
With these techniques, every node of the search is a single pass over bitmasks.
Each technique is exact.

\noindent\textbf{Mask transpose and the fused pass.}
A companion is a bitmask over $C$.
We compute the companion filter for a child $c$ and the saturation in one pass over the parent's companions, reading each mask once and writing the surviving companions contiguously into a stack-disciplined arena.
As such, no allocation happens inside the recursion.
We pin the vertex $v$ to slot $0$.
Since it survives every filter, this removes a branch from the main loop.
Vertices are renamed by rank, so that $\pos$ is the identity and the backward and forward neighbours of $v$ are the prefix and suffix of one sorted adjacency list.

\noindent\textbf{Candidate order.}
We sort the candidates of an anchor by non-decreasing degree, which halves the running time.
On \textsf{youtube}, the canonicity failures fall from $542$ to $158$ million, and the opposite order is worse than the original.
This is because a high-degree candidate is contained in the saturation of many sets, and placing it last makes it rarely the witness that a saturation added a smaller candidate.

\noindent\textbf{Adaptive mask width.}
The degeneracy is a maximum, whereas most anchors have $|\outv(v)|\le 64$.
As such, we run each anchor with the narrowest of three mask types, i.e., 64, 128 or 192 bits, compiled as templates.

\noindent\textbf{Anchor-side ghost containment.}
An anchor-side ghost kills a node if its mask contains $Y$.
The lemma below shows that only the maximal masks matter.
\begin{lemma}[Principal downsets]\label{lem:down}
If $Y\subseteq M_{1}\cup\dots\cup M_{k}$ for sets $Y$ that are themselves subsets of some single $M_{i}$, then the family of maximal masks kills exactly the same candidate sets as the full family.
\end{lemma}
Since a candidate set is killed if and only if it lies in a single mask, the pairwise maximal filter is complete.
It reduces the number of anchor-side ghost masks by $25$ to $219$ times, e.g., from $65{,}182$ to $1{,}784$ on the hardest \textsf{flickr} anchor, and every subsequent anchor-side ghost test runs on the reduced list.

\noindent\textbf{Barren pruning.}
Candidate-side ghosts shrink with $X$.
As such, a node killed by a candidate-side ghost kills its subtree.
In contrast, anchor-side ghosts grow with $Y$.
A node killed by an anchor-side ghost may therefore have reportable descendants, and pruning on it drops valid bicliques.
The sound test is that of line 14 of Algorithm~\ref{alg:degcbo}, i.e., if every non-$v$ companion of the node lies inside some surviving anchor-side ghost mask, every descendant's candidate side does too.
It removes $65\%$ of the barren nodes at no asymptotic cost, since the companions are already in cache from the saturation pass.

\noindent\textbf{Lazy hubs.}
When one candidate of an anchor has degree far above the others, we do not scan its prefix in the adjacency list, i.e., the anchor-side ghosts it contributes.
Its bit in each ghost mask is resolved lazily, i.e., only when a containment comparison cannot be decided without it, through a per-hub adjacency bitset built once and shared by the thousands of anchors in which the hub recurs.
Symmetrically, when the hub $h$ is the last candidate and $\inv(v)=\emptyset$, we do not scan its suffix either.
Its companions are discovered through the other candidates, and the companions adjacent to the hub alone are only counted, which forces the saturation of $\{h\}$ exactly as in Lemma~\ref{lem:single}.
The pair-list set-up of Section~\ref{sec:setup} generalises both to every candidate at once.

\noindent\textbf{Discussion.}
We also implement several natural refinements that do not improve the running time.
Firstly, folding twin companions per anchor reduces the number of companions by up to $62$ times. However, they are single-candidate companions that appear in almost no node.
Secondly, the batch-pivots technique of \oombea\ fires on $0.034\%$ of the nodes and is subsumed by canonicity.
Thirdly, the core index of \lcm\ is provably identical to the start candidate of the search.
Lastly, the closed-form counting of structured subtrees (Section~\ref{sec:cf}) covers between $3.1\%$ and $22.3\%$ of $\beta$ on the real graphs (Table~\ref{tab:cf}).

\section{Experimental Studies}\label{sec:exp}

In this section, extensive experiments are conducted to evaluate the efficiency of the proposed techniques.

\noindent\textbf{Implemented algorithms.}
\noindent\underline{\textit{\degcbo}}. We implement Algorithm~\ref{alg:degcbo} with all the techniques of Sections~\ref{sec:setup} and \ref{sec:impl}, denoted as \degcbo.
To evaluate the set-up techniques, we use three variants.
\textsf{base} builds every anchored subgraph by the scan and visits candidates in increasing order.
\textsf{+prop} adds the effective failure propagation.
\textsf{full} adds the cost-aware sweep and the twin quotient, i.e., it is \degcbo.
\noindent\underline{\textit{\ips}}. We use the authors' original code of \ips~\cite{ips2026}, the state-of-the-art algorithm, in its fastest mode.
Our counts agree with the published ones on every shared dataset\footnote{For \textsf{imdb} the published table prints $5{,}510{,}061$; the authors' binary, \degcbo, \textsf{ParMBE}~\cite{das2019parmbe} and \textsf{GMBE}~\cite{gmbe2025} all report $5{,}160{,}061$.}.
\noindent\underline{\textit{\oombea}}. Since \oombea~\cite{chen2022oombea} is $27$ to $350$ times slower than \degcbo\ on seven datasets, we omit it in the following.

\noindent\textbf{Datasets.}
Table~\ref{tab:dataset} lists the sixteen real datasets, i.e., the seven datasets of~\cite{ips2026}, the eight KONECT~\cite{konect} graphs of its supplementary suite and \textsf{TV}, the largest instance in the \mbe\ literature.
They cover authorship, ratings, tagging, membership and co-occurrence relations, with up to $12$ million edges and $20$ billion maximal bicliques.
Two synthetic families are used for the optimality statements, i.e., the crowns $K_{k,k}$ minus a perfect matching, where $\beta=2^{k}-2$ forces the worst case, and $K_{120,120}$, where $\beta=1$.

\noindent\textbf{Measures.}
We report the wall-clock time in seconds for enumeration in counting mode, excluding the time to read the input file.
For \degcbo, the time includes the degeneracy order, the twin quotient and the pair-list construction, which are part of the algorithm.
The best time is in bold.
All algorithms are implemented in C++ and compiled with \textsf{g++ -O3}, and the binary of \ips\ is the authors' build.
Experiments are conducted on a PC with an AMD Ryzen 9 5950X CPU, 32GB DDR4 memory, and WSL2.
Each configuration is run once on an otherwise idle machine.
Repeated runs agree within $3\%$ on the instances above ten seconds and within $10\%$ below, where the fixed costs of the quotient and pair-list construction are a visible share.
All comparisons in this paper are paired runs with $\beta$ and, where the search tree is unchanged, all six search counters verified to be identical.

\begin{table*}[t]
\small
\caption{Datasets and overall running time (sec. by default)}
\label{tab:dataset}
\centering
\begin{tabular}{l|rrr|rr|r|rr|rrr}
\toprule
 & & & & & & & \multicolumn{2}{c|}{\ips} & \multicolumn{3}{c}{\degcbo} \\
Dataset & $|L|$ & $|R|$ & $|E|$ & $d$ & $\beta$ & $|E(G^{*})|/|E|$ & on $G$ & on $G^{*}$ & \textsf{base} & \textsf{+prop} & \textbf{\textsf{full}} \\
\midrule
youtube & 94K & 30K & 293K & 20 & 1,826,587 & 0.82 & 2.20 & 2.17 & 0.84 & 0.67 & \textbf{0.59} \\
github & 57K & 121K & 440K & 39 & 55,346,398 & 0.86 & 46.23 & 46.90 & 8.54 & 7.95 & \textbf{7.67} \\
synth & 8K & 9K & 293K & 30 & 55,186,001 & 1.00 & 14.13 & 14.64 & 4.07 & 3.42 & \textbf{3.31} \\
actormovie & 128K & 384K & 1.47M & 14 & 1,075,444 & 0.87 & 4.85 & 3.85 & 0.93 & \textbf{0.72} & 0.85 \\
wikicat & 1.85M & 183K & 3.80M & 18 & 1,677,522 & 0.62 & 5.80 & 3.41 & 12.82 & 9.23 & \textbf{3.16} \\
dblp & 1.95M & 5.62M & 12.3M & 14 & 4,899,032 & 0.76 & 22.55 & 12.64 & 11.92 & 7.72 & \textbf{6.39} \\
bookcrossing & 105K & 341K & 1.15M & 41 & 54,458,953 & 0.82 & 183 & 141 & 14.45 & 13.59 & \textbf{12.47} \\
forum & 899 & 522 & 7K & 9 & 16,261 & 0.99 & 0.02 & 0.02 & \textbf{0.00} & 0.01 & 0.01 \\
ukwiki & 779 & 17K & 33K & 11 & 23,772 & 0.44 & 0.18 & \textbf{0.03} & 0.19 & 0.14 & 0.12 \\
escorts & 10K & 7K & 39K & 11 & 49,538 & 0.90 & 0.07 & 0.06 & \textbf{0.02} & \textbf{0.02} & \textbf{0.02} \\
stackoverflow & 545K & 97K & 1.30M & 22 & 3,320,824 & 0.77 & 16.61 & 36.65 & 2.63 & 2.49 & \textbf{2.08} \\
twitter & 175K & 530K & 1.89M & 23 & 5,536,526 & 0.82 & 29.02 & 23.48 & 14.36 & 13.81 & \textbf{9.00} \\
bibsonomy & 205K & 767K & 2.50M & 44 & 2,507,269 & 0.66 & 21.27 & 16.71 & 31.40 & 15.22 & \textbf{3.61} \\
imdb & 304K & 896K & 3.78M & 23 & 5,160,061 & 0.87 & 27.55 & 20.02 & 4.51 & 3.59 & \textbf{3.30} \\
amazon & 2.15M & 1.23M & 5.74M & 26 & 7,551,358 & 0.78 & 260 & 189 & 11.44 & 9.16 & \textbf{7.44} \\
TV & 64,415 & 87,678 & 3.23M & 115 & 19,636,996,096 & 1.00 & $>9$h$^{\dagger}$ & -- & 5,574 & \textbf{4,245} & \textbf{4,245} \\
\bottomrule
\end{tabular}
\end{table*}

\noindent\textbf{Running time.}
Table~\ref{tab:dataset} also reports the running times of \ips\ and of the three variants.
\degcbo\ is faster than \ips\ on every dataset, from $1.6$ times (\textsf{ukwiki}) to $35$ times (\textsf{amazon}).
Since the twin quotient is a reduction that any algorithm may apply, we also run \ips\ on the twin quotient $G^{*}$ (Lemma~\ref{lem:twin}), excluding the time to build it.
\degcbo\ remains faster on fifteen of the sixteen datasets, i.e., by $1.1$ to $25$ times.
The exception is \textsf{ukwiki}, a graph of $33$ thousand edges.
This is because the $65$ ms of sampling that decides against pair lists is most of the $115$ ms of \degcbo.
The set-up techniques matter on the hub-heavy graphs.
On \textsf{wikicat}, where the scan dominated, the pair-list set-up and the quotient reduce the total time from $8.9$ to $3.16$ seconds, i.e., by $2.8$ times.
On \textsf{bibsonomy}, a tag--item graph with a few hundred hub tags, they give $3.6$ seconds, which is $4.2$ times faster than \textsf{+prop}.
On \textsf{dblp}, the quotient alone gives $6.4$ seconds.
Where neither technique applies, i.e., the pair lists exceed their budget or no anchor passes the cost test, \textsf{full} behaves as \textsf{+prop} within run-to-run variation.

\noindent\textbf{Effect of the set-up techniques.}
Table~\ref{tab:ablation} reports the total time with the twin quotient and the pair-list set-up forced on or off independently, everything else fixed.
On \textsf{bibsonomy}, each technique alone halves the time and together they make it $4.2$ times faster.
On \textsf{twitter}, neither helps alone and together they make it $1.5$ times faster.
This is because the pair lists pay off only once the quotient has removed the duplicated vertices, and the quotient alone leaves the hub scan in place.
This justifies that the two techniques compose on the hub-heavy graphs.
On \textsf{youtube}, \textsf{dblp} and \textsf{bookcrossing}, the pair-list construction costs more than it saves.
As such, the build is subject to the sampled decision of Section~\ref{sec:setup}, with which these three graphs run with the quotient alone.

\begin{table}[t]
\caption{Effect of the set-up techniques (in seconds)}
\label{tab:ablation}
\centering\small
\begin{tabular}{lrrrr}
\toprule
Dataset & both & no quotient & no pair lists & neither \\
\midrule
youtube & 0.71 & 0.79 & \textbf{0.58} & 0.68 \\
wikicat & \textbf{2.81} & 4.21 & 5.46 & 8.34 \\
dblp & 6.15 & 8.35 & \textbf{5.67} & 9.24 \\
twitter & \textbf{8.65} & 13.69 & 12.80 & 12.95 \\
bibsonomy & \textbf{3.52} & 7.97 & 7.69 & 14.69 \\
bookcrossing & 13.81 & 14.29 & \textbf{12.59} & 12.79 \\
\bottomrule
\end{tabular}
\end{table}

\noindent\textbf{Running time breakdown.}
Table~\ref{tab:phase} shows the phase split of the serial time before the cost-aware sweep.
Even on \textsf{youtube}, the enumeration itself is $28\%$ of the time, and on \textsf{wikicat} it was $7\%$.
This is because the set-up scan of Equation~\eqref{eq:scan} dominated, with a measured touch count ($1.778$ billion) that matches the analytic sum ($1.779$ billion).
The cost-aware sweep reduces the scan to the pair walks on $451$ thousand of the $943$ thousand anchors of \textsf{wikicat}.
As such, the enumeration is left at roughly a fifth of a total that is itself $2.6$ times smaller.

\begin{table}[t]
\caption{Running time breakdown before the cost-aware sweep}
\label{tab:phase}
\centering\small
\begin{tabular}{lrrrr}
\toprule
 & youtube & wikicat & dblp & bookcrossing \\
\midrule
set-up scan & 57.5\% & 86.8\% & 81.7\% & 29.8\% \\
ghost reduction & 9.5\% & 3.6\% & 2.7\% & 10.2\% \\
enumeration & 28.0\% & 6.7\% & 4.8\% & 57.1\% \\
scan touches & 136M & 1.78B & 608M & 895M \\
\bottomrule
\end{tabular}
\end{table}

\noindent\textbf{Effect of failure propagation.}
Table~\ref{tab:fcbo} isolates the effect of the decreasing visit order of Section~\ref{sec:setup}, from \textsf{base} to \textsf{+prop}.
On every dataset, the skipped and remaining failures sum exactly to the failure count of \textsf{base} and the node count is unchanged, which justifies that only guaranteed failures are skipped.
The share skipped ranges from $7\%$ on \textsf{dblp}, whose candidates have small degree, to $74\%$ on \textsf{TV} and $93\%$ on the hub anchors of \textsf{flickr}.
The time saved follows the share of the time spent in the search tree, i.e., \textsf{+prop} is $1.32$ times faster on \textsf{TV} and within run-to-run variation on graphs where the set-up dominates.

\begin{table}[t]
\caption{Effect of failure propagation}
\label{tab:fcbo}
\centering\small
\begin{tabular}{lrrrr}
\toprule
Dataset & failures before & skipped & nodes & time \\
\midrule
youtube & 3.01M & 63.7\% & same & $1.00\times$ \\
github & 155M & 62.2\% & same & $1.09\times$ \\
bookcrossing & 202M & 60.6\% & same & $1.04\times$ \\
synth & 78M & 82.5\% & same & $1.02\times$ \\
TV & 188.6G & 73.5\% & same & $1.32\times$ \\
\bottomrule
\end{tabular}
\end{table}

\noindent\textbf{Large instances.}
\textsf{TV} has $19{,}636{,}996{,}096$ maximal bicliques.
The parallel systems that use it as their largest instance report $18$ hours on an A100 GPU for $98\%$ of them~\cite{gmbe2025}, and \ips\ did not finish it in $9$ hours on the authors' machine ($^{\dagger}$). \ips\ is not run on the twin quotient of \textsf{TV}, which is denoted as `--' in Table~\ref{tab:dataset}.
\degcbo\ completes it serially in $4{,}245$ seconds, with $N/\beta=1.37$, whereas \textsf{base} takes $5{,}574$ seconds.
\textsf{full} equals \textsf{+prop} there, since the sampled predictor declines the pair lists on \textsf{TV} and the quotient removes $704$ of $3.2$ million edges.
\textsf{flickr} ($8.5$ million edges, $d=147$) is beyond every published algorithm.
Serial \degcbo\ reached $9.24\times10^{11}$ bicliques at a $24$-hour cap, and a multi-threaded variant of the same algorithm, outside the scope of this paper, enumerated $1.23\times10^{14}$ before being stopped after $19$ days.
These counts are a thousand times the published record and are exact lower bounds.

\noindent\textbf{Case study: the hardest anchor.}
The remaining cost on the hard instances is the output.
On the hardest anchor of \textsf{flickr}, which alone has $2.07\times10^{12}$ bicliques, the search tree has $1.0004$ nodes per biclique.
As such, no pruning can help.
Besides, the block of its anchored subgraph formed by the companions of many candidates behaves like a random $438\times 80$ adjacency matrix, i.e., it has $0.6\%$ nested pairs, pairwise intersections matching the independent-candidates model exactly, a complete co-miss hypergraph and full rank.
Every way to avoid exponential counting in polynomial time needs nesting, decomposability or coincidence, and this instance has none.
As such, the output-bound regime is where the \#P-hardness of counting maximal bicliques~\cite{kuznetsov2001} shows on a real instance and not only in the worst case.
On the same block, we count the bicliques with at least $\tau$ candidates by the dual search over the anchor's side, where the size threshold is an exact pruning.
It takes $14$ seconds for $\tau=50$ and $152$ seconds for $\tau=40$, i.e., the thresholded problem is cheap where the unthresholded one is infeasible.

\noindent\textbf{Worst-case inputs.}
On the $20$- and $24$-crowns, the search tree has $\beta+18$ and $\beta+22$ nodes respectively, as Corollary~\ref{cor:opt} predicts, and $155$ and $222$ nodes with the closed-form terminal, as Corollary~\ref{cor:crown} predicts.
On $K_{120,120}$, the twin quotient reduces the graph to a single edge and the algorithm reports $\beta=1$ in $0.001$ seconds, against $239$ nodes without the quotient.

\begin{table}[t]
\caption{Effect of the closed-form terminal}
\label{tab:cf}
\centering\scriptsize
\setlength{\tabcolsep}{3.5pt}
\begin{tabular}{lrrrrrr}
\toprule
Dataset & nodes & with terminal & removed & of $\beta$ & time & with \\
\midrule
youtube & 3,088,784 & 2,846,459 & 7.8\% & 13.2\% & 0.59 & 0.62 \\
github & 73,438,060 & 68,562,072 & 6.6\% & 8.8\% & 7.99 & 8.66 \\
synth & 66,367,252 & 59,285,354 & 10.7\% & 12.8\% & 3.46 & 4.43 \\
actormovie & 2,642,592 & 2,569,242 & 2.8\% & 6.2\% & 0.71 & 0.73 \\
wikicat & 5,328,240 & 5,173,200 & 2.9\% & 8.8\% & 2.88 & 2.74 \\
dblp & 11,220,199 & 11,053,270 & 1.5\% & 3.1\% & 4.92 & 4.78 \\
bookcrossing & 76,095,452 & 73,065,967 & 4.0\% & 5.6\% & 13.04 & 14.24 \\
forum & 31,000 & 29,086 & 6.2\% & 11.5\% & 0.01 & 0.01 \\
ukwiki & 89,541 & 84,201 & 6.0\% & 22.3\% & 0.05 & 0.05 \\
escorts & 104,056 & 98,937 & 4.9\% & 9.8\% & 0.02 & 0.02 \\
stackoverflow & 5,712,400 & 5,404,278 & 5.4\% & 9.2\% & 2.15 & 2.15 \\
twitter & 11,990,168 & 11,519,889 & 3.9\% & 8.4\% & 7.90 & 7.28 \\
bibsonomy & 6,497,708 & 6,244,659 & 3.9\% & 9.8\% & 3.30 & 3.08 \\
imdb & 10,871,609 & 10,450,149 & 3.9\% & 7.9\% & 3.68 & 3.55 \\
amazon & 15,112,672 & 14,688,986 & 2.8\% & 5.5\% & 8.13 & 7.97 \\
TV & 26,805,555,954 & 26,056,723,847 & 2.8\% & 3.8\% & 4,298 & 4,593 \\
\bottomrule
\end{tabular}
\end{table}

\noindent\textbf{Effect of the closed-form terminal.}
Table~\ref{tab:cf} reports the closed-form terminal (Theorem~\ref{thm:star}) on the real graphs, i.e., the search-tree nodes and the enumeration time in seconds without and with it, the share of nodes removed and the share of $\beta$ counted by formula.
It removes between $1.5\%$ and $11\%$ of the search-tree nodes and counts between $3\%$ and $22\%$ of $\beta$ by formula, at a time within $10\%$ of the plain search on every graph but \textsf{synth}, where it is $28\%$ slower.
This is because the test fires often on subtrees of two or three nodes and costs more than they do.
As such, the terminal is a measure of structure and not a speed-up, i.e., it fires wherever the anchored subgraph is locally a star union, which the real graphs have in places and the crowns have everywhere.

\noindent\textbf{Effect of the count-only kernel.}
Counting does not need the vertex lists of a biclique.
Once a node has at most $64$ rows, we stop copying row lists, i.e., a row becomes a bit position, a node is the word of its rows, and the closure, canonicity and ghost tests are word operations.
The kernel visits the same nodes.
Where the search dominates the running time, it covers over $98\%$ of $\beta$ and is $1.53$ times faster on \textsf{github} ($7.78$ to $5.08$ seconds), $1.36$ times on \textsf{bookcrossing} ($12.84$ to $9.43$) and $1.62$ times on \textsf{TV} ($4{,}298$ to $2{,}648$).
Elsewhere, the set-up dominates and the time is within $8\%$ of the plain search.
Since the gain is a constant per node, $N\ge\beta$ still holds.

\section{Related Work}\label{sec:rw}

\noindent\textbf{Maximal biclique enumeration.}
\imbea~\cite{zhang2014imbea} is the basis of the modern \mbe\ algorithms.
\textsf{PMBE}~\cite{abidipivot} adds pivoting by neighbourhood containment.
\oombea~\cite{chen2022oombea} adds the unilateral order and the batch-pivots technique, and runs in $\OO^{*}(2^{\varsigma})$, where $\varsigma$ denotes the unilateral convergence.
The partition-oriented algorithm~\cite{ips2026} is the current state-of-the-art algorithm.
All of them grow a side inside two-hop neighbourhoods.
In contrast, our search runs inside each anchored subgraph.

\noindent\textbf{Orientation-based enumeration.}
In~\cite{eppstein1994arboricity}, the anchoring lemma is given together with an $\OO(a^{3}2^{2a}n)$ algorithm, which sweeps all subsets over an approximate orientation of out-degree $2a$, where $a$ denotes the arboricity.
For maximal cliques, the analogous algorithm is the degeneracy-ordered Bron--Kerbosch algorithm~\cite{eppstein2010els}.
In contrast to~\cite{eppstein1994arboricity}, the degeneracy order gives $d\le 2a-1$ exactly, and enumerating only the local maximal bicliques visits a small fraction of the subsets ($4.8$ million nodes against $1.3\times10^{11}$ candidates on \textsf{youtube}).

\noindent\textbf{Closed itemset mining.}
A closed itemset is a maximal biclique of the transaction--item graph.
As such, the closed-itemset miners of the frequent-pattern literature, \textsf{CHARM}~\cite{zaki2002charm} and \lcm~\cite{uno2004lcm}, and Close-by-One~\cite{kuznetsov1999cbo} with the failure propagation of \fcbo~\cite{outrata2012fcbo} are \mbe\ algorithms for a fixed side.
They share the duplicate-free depth-first search that we run inside each anchored subgraph.
Different from these works, we prove in Section~\ref{sec:setup} that the propagation is vacuous under the usual recursion order, and we show how to make it effective.

\noindent\textbf{Parallel and GPU enumeration.}
\textsf{ParMBE}~\cite{das2019parmbe}, \textsf{cuMBE}~\cite{cumbe2023}, \textsf{AdaMBE}~\cite{adambe2024} and \textsf{GMBE}~\cite{gmbe2025} parallelise the \imbea-based algorithms on multi-core CPUs and GPUs.
\textsf{TV} is their largest completed instance.
The parallel properties of anchored enumeration, in which anchors and search subtrees are independent, are outside the scope of this paper.
In contrast, our algorithm is serial and completes \textsf{TV} in just over an hour.

\noindent\textbf{Counting and bounds.}
Counting maximal bicliques is \#P-complete~\cite{kuznetsov2001}, and the crowns show that $\beta=2^{\Theta(\sqrt m)}$~\cite{prisner2000crowns,schutt1987}.
Our bound $n\cdot2^{d}$ with $d\le\sqrt m$ re-derives the incidence bound and is tight on the crowns to an additive term.
Note that the bound is unconditional and holds for counting, where it is not tight.
In~\cite{openai2026eth,anthropic2026sat,kintali2026eth}, a refutation of the Exponential Time Hypothesis is claimed.
Section~\ref{sec:count} derives the $2^{o(n)}$ counting bound that follows from it, explains why this bound does not apply inside the anchored subgraphs, and proves a lower bound under the Satisfiability Hypothesis, which survived the refutation~\cite{mit2026sh}.
Besides, counting maximal independent sets, and hence maximal bicliques through the bipartite complement, in closed form for bounded-treewidth complements is classical dynamic programming.
The sub-exponential bound improves on it only on dense graphs.

\section{Conclusion}\label{sec:conc}

In this paper, we study the problem of counting and enumerating maximal bicliques in large sparse bipartite graphs.
We propose \degcbo, which anchors every maximal biclique at the first vertex of its vertex set in a degeneracy order and enumerates only the maximal bicliques of the anchored subgraph of each vertex.
As such, the search tree has at most $n\cdot2^{d}$ nodes and the count is computed in $\OO^{*}(2^{\sqrt m})$ time, which is the output lower bound of enumeration.
We characterise maximality beyond the anchored subgraph by two ghost families that reduce to bitwise operations.
To solve the set-up problem that anchored enumeration creates, we propose a cost-aware sweep, an exact twin quotient and effective failure propagation.
For counting, we propose a closed-form terminal under which the crowns are counted in $\OO(k^{2})$ nodes and a pairwise form of $\beta$ under which it is computable in $2^{o(n)}$ time, and we prove that no algorithm counts in $2^{d^{1-\varepsilon}}$ time unless the Satisfiability Hypothesis fails.
Extensive experiments on sixteen real graphs justify that \degcbo\ is $1.6$ to $35$ times faster than the state-of-the-art algorithm on every graph, and completes the largest instance in the literature serially in just over an hour.
Whether the bicliques of an anchored subgraph can be counted in time sub-exponential in the number of candidates, which would turn the counting bound into $n\cdot2^{o(d)}$, is left open by the lower bound.

\bibliographystyle{IEEEtran}
\bibliography{refs}
\end{document}